\documentclass[12pt]{iopart}

\usepackage{iopams} 
\usepackage{graphicx}
\usepackage{graphics}
\usepackage{subfigure}
\usepackage{bm}
\usepackage{float}
\usepackage{tabu}
\usepackage{xcolor}
\usepackage{color}
\usepackage{makecell}
\usepackage{blkarray, bigstrut}

\begin{document}
\title[Unbending strategies shepherd cooperation]{Unbending strategies shepherd cooperation and suppress extortion in spatial populations}

\author{Zijie Chen$^{1}$, Yuxin Geng$^{1}$, Xingru Chen$^{1}$ and Feng Fu$^{2,3}$}

\address{$^1$School of Science, Beijing University of Posts and Telecommunications, Beijing 100876, China}
\address{$^2$Department of Mathematics, Dartmouth College, Hanover, NH 03755, USA}
\address{$^3$Department of Biomedical Data Science, Geisel School of Medicine at Dartmouth, Lebanon, NH 03756, USA}
\ead{xingrucz@gmail.com}
%\ead{fufeng@gmail.com}

\begin{abstract}
Evolutionary game dynamics on networks typically consider the competition among simple strategies such as cooperation and defection in the Prisoner's Dilemma and summarize the effect of population structure as network reciprocity. However, it remains largely unknown regarding the evolutionary dynamics involving multiple powerful strategies typically considered in repeated games, such as the zero-determinant (ZD) strategies that are able to enforce a linear payoff relationship between them and their co-players. Here, we consider the evolutionary dynamics of always cooperate (AllC), extortionate ZD (extortioners), and unbending players in lattice populations based on the commonly used death-birth updating. Out of the class of unbending strategies, we consider a particular candidate, PSO Gambler, a machine-learning-optimized memory-one strategy, which can foster reciprocal cooperation and fairness among extortionate players. We derive analytical results under weak selection and rare mutations, including pairwise fixation probabilities and long-term frequencies of strategies. In the absence of the third unbending type, extortioners can achieve a half-half split in equilibrium with unconditional cooperators for sufficiently large extortion factors. However, the presence of unbending players fundamentally changes the dynamics and tilts the system to favor unbending cooperation. Most surprisingly, extortioners cannot dominate at all regardless of how large their extortion factor is, and the long-term frequency of unbending players is maintained almost as a constant. Our analytical method is applicable to studying the evolutionary dynamics of multiple strategies in structured populations. Our work provides insights into the interplay between network reciprocity and direct reciprocity, revealing the role of unbending strategies in enforcing fairness and suppressing extortion.
\end{abstract}

\vspace{2pc}

\submitto{\NJP}

\maketitle
% 
% For two-column output uncomment the next line and choose [10pt] rather than [12pt] in the \documentclass declaration
%\ioptwocol
%

\section*{Introduction}

A central aspect of human behavior is cooperation~\cite{perc2017statistical}. Cooperation is needed for collective action problems, ranging from climate change~\cite{wang2009emergence, vasconcelos2014climate} to pandemic control~\cite{glaubitz2024social}. However, cooperation incurs a cost to oneself and benefits others; therefore it can be vulnerable to exploitation by defection, which pays no cost and free rides on others' effort~\cite{hauert2005game}. The Prisoner's Dilemma game is commonly used to illustrate this tension~\cite{axelrod1981evolution}. Two individuals play this game simultaneously. If both cooperate, they each receive a reward, $R$, for mutual cooperation. If one cooperates while the other defects, the cooperator receives the sucker's payoff, $S$, while the defector gets the payoff of the temptation to defect, $T$. If both defect, they both receive the punishment of mutual defection, $P$. We have $T >  R > P > S$ in order for the game to be classified as a Prisoner's Dilemma~\cite{rapoport1989prisoner}.

To resolve the conundrum of cooperation, different mechanisms have been extensively studied~\cite{nowak2006five}. One important mechanism, called direct reciprocity, is to repeat the game and investigate how individuals can build mutually beneficial cooperation when engaged in multiple encounters. This extension leads to the famous framework of the Iterated Prisoner's Dilemma (IPD) with many discoveries worthy of mention. Among others, fair-minded Tit-for-Tat (`I will if you will')~\cite{nowak1992tit} and adaptive `Win-Stay, Lose-Shift' (keep using the current strategy after payoff outcomes $R$ and $T$ but switch otherwise)~\cite{nowak1993strategy} are notable. In noisy IPD games, generous Tit-for-Tat prevails over unforgiving ones~\cite{nowak1992tit, molander1985optimal}. Moreover, equalizer strategies are able to unilaterally set any co-player's payoff to a given value within the interval $[P, R]$, therefore the name equalizer~\cite{boerlijst1997equal}. More general than equalizers, Press and Dyson have discovered the so-called zero-determinant (ZD) strategies, which are able to enforce a linear payoff relation between themselves and their co-players~\cite{press2012iterated}. A subclass of ZD strategies, called extortionate ZD, is parameterized by the baseline payoff $P$ and the extortion factor $\chi  >  1$ such that the two payoffs $\pi_X$ and $\pi_Y$ of a ZD player X playing against an opponent Y satisfy $\pi_X - P = \chi (\pi_Y - P)$. Extortionate ZD seems formidable but is revealed to have the Achilles' heel~\cite{chen2023outlearning}: its dominance and extortion ability are impacted by the underlying payoff structure and under $T+S < 2P$ it can actually have a lower payoff than Win-Stay, Lose-Shift. Inspired by the empirical results~\cite{hilbe2014extortion}, a recent theoretical study uncovers unbending strategies where extortion does not pay off and leads to lower payoffs than being more fair~\cite{chen2023outlearning}. 

Another important mechanism, called network or spatial reciprocity, is to consider structured populations where individuals' interactions are not random but exquisitely patterned as a graph or a network, such as the square lattice~\cite{nowak1992evolutionary, lindgren1994evolutionary, perc2006coherence}. The seminal work of spatial games has ushered in an era of studying games on networks~\cite{hauert2004spatial, santos2005scale, ohtsuki2006simple, poncela2007robustness, szolnoki2008coevolution, fu2009evolutionary, jackson2015games, su2016interactive, perez2022cooperation, wu2023evolutionary}. Broadly speaking, population structure leads to assortment, meaning like-with-alike, in a way that clusters of cooperators can resist invasion by neighboring defectors and reaching a dynamic balance~\cite{nowak1992evolutionary, wu2023evolutionary}. In the donation game (a simplified Prisoner's Dilemma) where a cooperator pays a cost $c$ for another to receive a benefit $b$ while a defector does nothing, the payoff matrix becomes $R = b - c$, $S = -c$, $T = b$, and $P = 0$. Prior studies have quantified how network structure, in particular the average degree $k$, impacts the evolution of cooperation, requiring $b/c > k$~\cite{ohtsuki2006simple}. 

This celebrated result for the evolution of cooperation on networks states that cooperators are able to prevail and be favored over defectors because of the network clustering effect in structured populations. More generally, the impact of population structure on strategy competition can be written as a more general form: $\sigma R + S > T + \sigma P$, which says that cooperators are more abundant than defectors if this condition holds with the coefficient $\sigma$ summarizing the effect of population structure~\cite{tarnita2009strategy, allen2017evolutionary}. Extending this formula to matrix games with multiple strategies yields similar conditions that can be obtained under the mutation-selection equilibrium~\cite{antal2009mutation, tarnita2011multiple, mcavoy2021fixation}. Besides these analytical insights, there has been significant interest in studying games on networks from the perspective of statistical physics~\cite{hauert2005game}, with numerous contributions from the field (for a recent review, for example, see Ref.~\cite{jusup2022social}).

Beyond simple strategies, a surge of curiosity has arisen in exploring the evolutionary dynamics of prescribed IPD strategies in well-mixed~\cite{hilbe2013evolution, stewart2013extortion} and structured populations~\cite{szolnoki2014evolution, szolnoki2014defection, wu2014boosting, xu2017extortion, mao2018emergence}. Part of these efforts is to study extortionate ZD from an evolutionary perspective~\cite{hilbe2013evolution, stewart2013extortion, hilbe2013adaptive, chen2022intricate}. The intuition that the lack of mutual cooperation among extortionate ZD makes them not evolutionary favorable. However, prior work has shown the impact of population size on the evolutionary advantage of extortionate ZD~\cite{hilbe2013evolution}, including some other general classifications of IPD strategies~\cite{akin2016iterated, hilbe2018partners, chen2023identifying}. Interestingly, previous work found that extortionate ZD can be a catalyst for cooperation~\cite{hilbe2013evolution, szolnoki2014evolution}. And simulations of a handful of IPD strategies show that this holds both in well-mixed populations and on networks~\cite{szolnoki2014defection, wu2014boosting, xu2017extortion, mao2018emergence}. Despite much progress made in this direction, it remains largely unknown how the resilience of cooperation can be bolstered in the wake of increased extortion factors by extortionate ZD aimed at securing even more inequality. Understanding the potential interplay between spatial reciprocity and direct reciprocity is especially important in light of recently discovered unbending strategies.

To address this issue, we use the IPD game with conventional payoff values (namely, $R = 3$, $S = 0$, $T = 5$, $P =1$), the same as in the well-known Axelrod tournament. Aside from always cooperate (AllC) and extortionate ZD typically considered in previous studies, we also include a third memory-one strategy called PSO Gambler. This strategy has been optimized using the machine learning algorithm particle swarm optimization (PSO)~\cite{harper2017reinforcement}. Notably, the PSO Gambler is found to have the \emph{unbending} property~\cite{chen2023outlearning}: the best response of extortionate ZD when playing against a fixed unbending player is to offer a fair split by letting the extortion factor be one. In other words, the larger the extortion factor, the greater the payoff reduction compared to what it would have been otherwise. Although we focus on these three particular strategies, our method works for any IPD strategy and can be extended to study more than three strategies in structured populations. We find that the presence of unbending individuals can greatly enhance the resilience of cooperation in a synergistic way that promotes direct reciprocity together with spatial reciprocity. In particular, the long-term frequency of unbending PSO Gamblers remains almost constant, being able to mitigate the negative impact enforced by extortionate ZD with increased extortion factors. 

\section*{Model and Methods}

\begin{figure*}[htbp]
\centering
 \includegraphics[width=0.9\columnwidth]{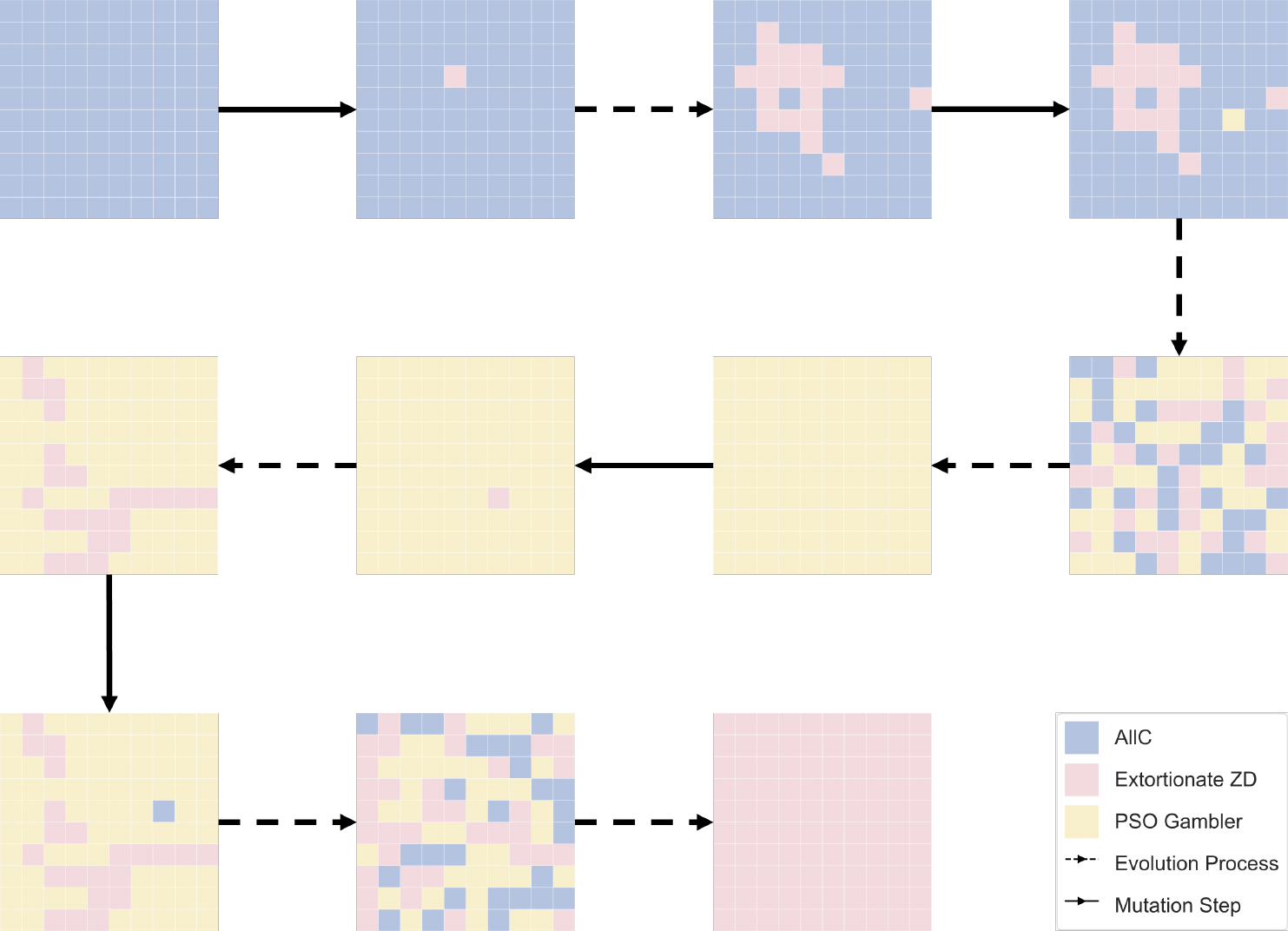}
        \caption{Spatiotemporal coevolutionary dynamics of AllC (always cooperate), extortionate zero-determinant strategy (ZD extortioner), and unbending strategy (PSO Gambler) in repeated games. The sequential series of spatial snapshots shows the invasion of a spatial population by mutations (indicated by solid arrows) and competition dynamics under the death-birth update rule (indicated by dashed arrows). The snapshots are taken from a stochastic simulation on a square lattice of size $10\times 10$ with von Neumann neighborhood $k = 4$, selection strength $\beta=0.001$, mutation rate $\mu=0.005$, conventional payoff parameters $R = 3$, $S = 0$, $T = 5$, $P =1$, $\chi =1$ for the zero-determinant strategy, and $[q_1, q_2, q_3, q_4] = [1, 0.52173487, 0, 0.12050939]$ for the memory-one PSO Gambler.}
        \label{fig1}
\end{figure*}

Here, we consider the evolutionary game dynamics of multiple strategies in spatial populations. To focus specifically on the interplay between network reciprocity and direct reciprocity, we consider strategies commonly used in the Iterated Prisoner Dilemma games, including always cooperate (AllC), extortionate zero-determinant (ZD) strategy (which is able to enforce an unfair split of payoffs with co-players, extortioners), and a particular candidate out of the unbending strategies (which can cause the backfire of extortion, PSO Gambler). The memory-one PSO Gambler used in our study is $\mathbf{q}_{\rm{PSO}} = [1, 0.52173487, 0, 0.12050939]$~\cite{harper2017reinforcement}. We consider a square lattice with the von Neumann neighborhood (degree $k = 4$) and periodic boundary conditions. Each node is occupied by an individual who adopts one of these three aforementioned strategies. An individual $i$ interacts with their immediate neighbors and accrues payoffs from their pairwise interactions, $\pi_i$. We use the exponential fitness function as $f_i = \exp(\beta \pi_i)$, where $\beta$ is the intensity of selection.

To account for the success of these three strategies, we use the following $3 \times 3$ expected payoff matrix to characterize their repeated interactions:
\begin{equation}
\begin{blockarray}{cccc}
& \rm{AllC} & \rm{ZD} & \rm{PSO} \\
\begin{block}{c[ccc]}
\rm{AllC}\,\,\, & a_{11} & a_{12} & a_{13} \\
\rm{ZD}\,\,\, & a_{21} & a_{22} & a_{23} \\
\rm{PSO}\,\,\, & a_{31} & a_{32} & a_{33} \\
\end{block}
\end{blockarray}
\,\,\, .
\end{equation}

Using the method originally introduced by Press and Dyson~\cite{press2012iterated}, these payoff entries can be analytically calculated by quotients of two determinants (see Appendix).

Specifically, for AllC versus extortionate ZD with a given extortion factor $\chi$, we have
\begin{equation}
\eqalign{
a_{11} &= R,\quad a_{12} = P + \frac{(R - P)(T - S)}{(R - S)\chi + (T - R)},\\
a_{22} &= P, \quad a_{21} = P + \frac{(R - P)(T - S)\chi}{(R - S)\chi + (T - R)}.
}
\label{abcpayoff}
\end{equation}
Moreover, the average payoff of extortion ZD against PSO Gambler, $a_{23}$, can be written as $a_{23} = g(\chi)/h(\chi)$, where both $g(\cdot)$ and $h(\cdot)$ are quadratic functions of $\chi$ and $g(\chi)/h(\chi)$ is monotonically decreasing for $\chi > 1$ with the maximum value of $R$ at $\chi =1$~\cite{chen2023outlearning}.

As for evolutionary updating, we consider death-birth updating with mutation. At each time step, an individual is randomly chosen to die and its neighbors compete for this vacant site with probability proportional to their fitness. A mutation occurs with probability $\mu$. The newly produced offspring is identical to its parent with probability $1 - \mu$; otherwise with probability $\mu$, it randomly chooses one of the three strategies (see Fig.~1). This process can also be interpreted in a cultural evolution setting with social imitation and exploration rates.

We perform stochastic agent-based simulations with asynchronous updating and average the frequencies of strategies over $1\times 10^7$ times. On top of that, closed-form predictions are feasible under the weak selection limit and with rare mutations. In the limit of low mutations, the fate of a new mutant is determined, either reaching fixation or going extinct, before the next mutant arises. Therefore, the population spends most of the time in homogeneous states with transitions in between given by the pairwise fixation probabilities, $\rho_{ij}$, for $1\le i, j \le 3$ and $i \neq j$: the probability that a single mutant of strategy $j$ invades and takes over the entire population of strategy $i$~\cite{ohtsuki2006simple}. The long-term frequencies of strategies can be approximated by the stationary distribution of the corresponding embedded Markov chain with the following transition matrix~\cite{fudenberg2006imitation, wu2012small, mcavoy2015comment}:
\begin{equation}
\mathbf{\Lambda} = \left(
\begin{array}{ccc}
1 - \mu(\rho_{12} + \rho_{13}) & \mu\rho_{12}  & \mu\rho_{13}  \\
\mu \rho_{21} & 1- \mu(\rho_{21}+\rho_{23})  & \mu\rho_{23}  \\
\mu \rho_{31} & \mu \rho_{32}  & 1 - \mu(\rho_{31}+\rho_{32})  \\
\end{array}
\right)
.
\end{equation}

\section*{Results}

We run agent-based simulations of the spatial system with the three aforementioned IPD strategies and are interested in their long-term frequencies. As depicted in Fig.~1, the spatial snapshots show the rise and fall of the respective three strategies and spatial population structures promote clustering (`like-with-alike') as foreseen. From time to time, mutations arise in the population, following which the mutants can form clusters and are likely to succeed in invading and taking over the system. Accordingly, Fig~1 shows possible transitions among homogeneous population states from AllC to PSO Gambler to extortionate ZD, along with the evolutionary dynamics of multiple strategies. Most of the time, evolutionary dynamics involve pairwise competition, but occasionally, all three strategies are present due to mutations, in particular as a consequence of neutral drift when $\chi = 1$.

\begin{figure*}[htbp]
\centering
 \includegraphics[width=0.9\columnwidth]{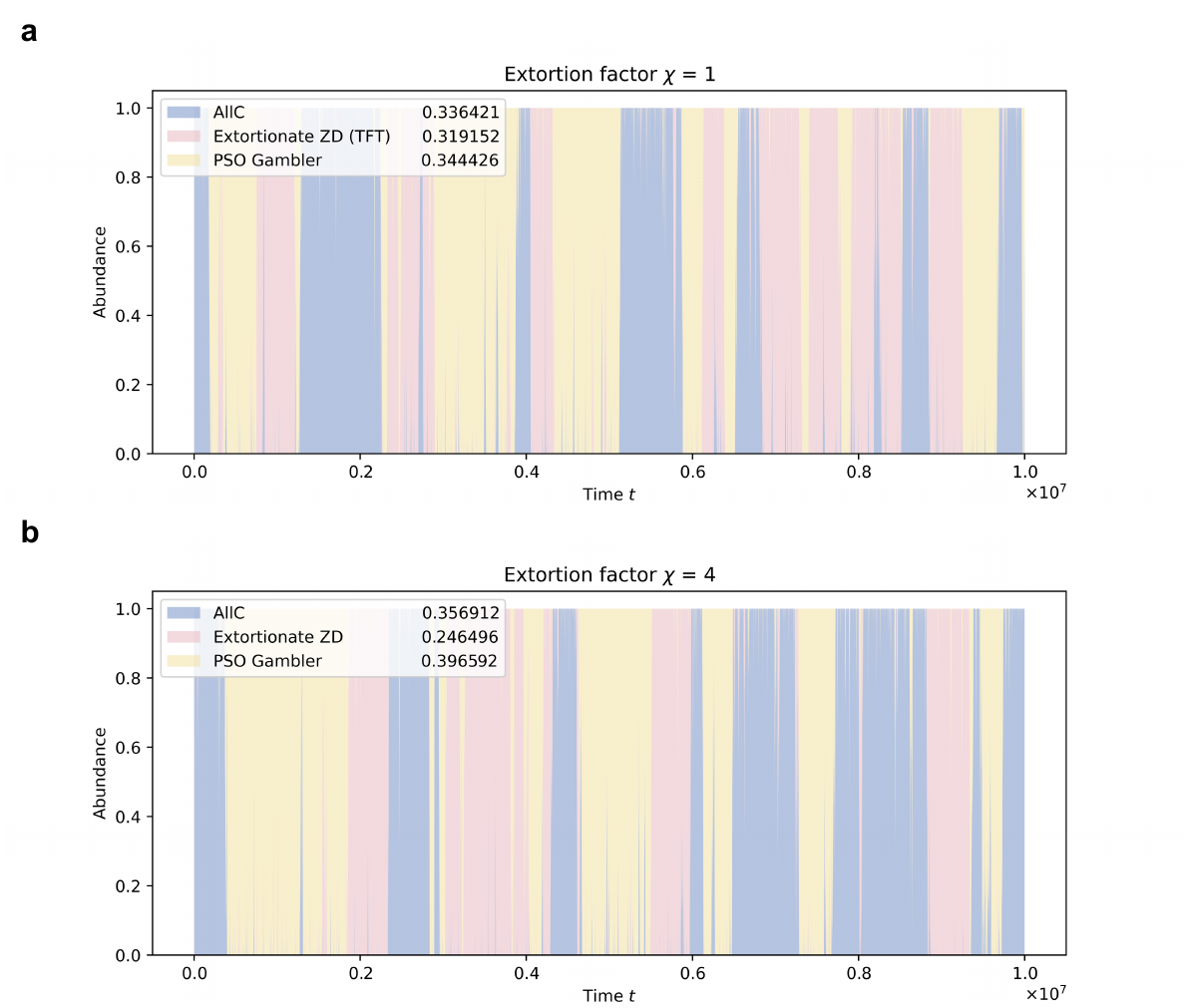}
        \caption{Time evolution of spatial competition dynamics among three strategies in repeated games: AllC, extortionate ZD, and PSO Gambler under rare mutations. Two different extortion factors for the extortionate ZD strategy are considered: (a) $\chi = 1$ which leads the extortioner to a Tit-for-Tat player, and (b) $\chi = 4$ which enables the extortioner X to unilaterally enforce an unfair payoff relation against its co-player Y as $s_X - P = \chi (s_Y - P)$. As expected, the simulation from (a) indicates neutral drift with equal frequencies of the three strategies. In contrast, the simulation from (b) suggests that the unbending strategy PSO Gambler is most abundant, and its presence can suppress unfair extortion with $\chi > 1$ and promote cooperation in spatial populations. Simulation parameters are as in Fig.~1, except that we also consider $\chi = 4$ for the extortionate ZD strategy.}
        \label{fig2}
\end{figure*} 
%[$\chi = 4$, cf.  panel (a)]

As a base case for comparison, we consider the special limiting case where the extortion factor $\chi = 1$ for the extortionate ZD strategy, which becomes the well-known Tit-for-Tat strategy~\cite{press2012iterated}, with the memory-one strategy representation as $\mathbf{q}_{\rm{TFT}} = [1, 0, 1, 0]$. This, in fact, leads to neutral dynamics among the three strategies considered since they each receive an equal payoff of mutual cooperation $R = 3$. As such, the long-term abundance of the three strategies should be equal to $1/3$. It can also be observed from Fig.~2a that our stochastic simulations confirm neutral drift dynamics, with the three strategies being roughly equally present in the population. 

In contrast, for an extortion factor $\chi > 1$, the extortionate ZD strategy is able to ensure a higher payoff (or at least equal) than any other strategy in a Prisoner's Dilemma satisfying $T + S > 2P$~\cite{chen2023outlearning}, which is why it is called extortioners in prior work~\cite{hilbe2013evolution}. The resulting evolutionary dynamics is thereby no longer neutral (compared with Fig.~2a). The extortion factor $\chi$ has a dual impact on extortioners: increasing $\chi$ makes them more fiercely aggressive against AllC players, but on the other hand, it reduces their absolute payoffs against PSO Gamblers. This stems from the PSO Gambler strategy being demonstrated to have the unbending property: against a fixed unbending player, any extortioner who intends to demand an unfairer payoff by increasing the extortion factor will lead to a lower payoff. As a consequence, unbending individuals are able to turn the tables against extortioners, suppress extortion, and shepherd cooperation. As shown in Fig.~2b, the unbending strategy, PSO Gambler, emerges as the most abundant in the long run, trailed by AllC, both are above $1/3$, while extortioners are suppressed and its abundance is below $1/3$.

In order to understand the underlying evolutionary dynamics in Fig.~2, we now consider pairwise competition dynamics that arise under rare mutations. In this limit, the dynamics can be studied by the stochastic transitions between homogenous states. The transitions are determined by the fixation probabilities $\rho_{ij}$ (see Model and Methods and also the Appendix for technical details).

Under weak selection, the pairwise fixation probability $\rho_{ij}$, the likelihood of a single individual of strategy $j$ taking over a spatial population of strategy $i$, has a closed-form expression up to the first order of $\beta$:
\begin{equation}
\fl
\eqalign{
\rho_{ij} &= \frac{1}{N} + \frac{(\ast)}{6(k-1)}\beta  + \mathcal{O}(\beta^2), \\
(\ast) &= (k+1)^2 a_{jj}  + (2k^2 - 2k -1)a_{ji} - (k^2 - k +1)a_{ij} - (2k-1)(k+1)a_{ii},
}
\label{fixprob}
\end{equation}
%O to o
where $N$ is the total population size, and $k$ is the degree of the lattice. This formula is derived using the discrete random walk approach as Ref.~\cite{fu2009evolutionary} (see Appendix), but it can also be obtained by the diffusion approximation method~\cite{ohtsuki2006simple}.

In the limit of low mutation (and weak selection), the long-term frequencies of strategies, $\lambda_i$, can be approximated using the embedded Markov chain approach and are related to the normalized left eigenvector of the transition matrix corresponding to the largest eigenvalue one: 
\begin{equation}
[\lambda_1, \lambda_2, \lambda_3] =\frac{1}{\gamma_1 + \gamma_2 +\gamma_3}[\gamma_1, \gamma_2, \gamma_3],
\end{equation}
where 
\begin{equation}
\eqalign{
\gamma_1 &= \rho_{21}\rho_{31} + \rho_{21}\rho_{32} + \rho_{31}\rho_{23}, \\
\gamma_2 &= \rho_{31}\rho_{12} + \rho_{12}\rho_{32} + \rho_{32}\rho_{13}, \\
\gamma_3 &= \rho_{21}\rho_{13} + \rho_{12}\rho_{23} + \rho_{13}\rho_{23}.
}
\end{equation}
Under weak selection ($\beta \to 0$), the condition for the stationary distribution of strategy $i$, $\lambda_i$, to be greater than 1/3 can be expressed equivalently as
\begin{equation}
\sum_{j = 1}^{3} (\rho_{ji} -  \rho_{ij}) > 0.
\end{equation}
Intuitively, this inequality indicates that the influx into strategy $i$ must exceed the outflux from it in the embedded Markov chain: $\sum_j{\rho_{ji}} >  \sum_j{\rho_{ij}}$. 

After substituting the fixation probabilities under weak selection in Eq.~\ref{fixprob} and simplifying the algebra, we obtain the following inequality for $\lambda_i >1/3$, namely, natural selection favors strategy $i$ and its long-run abundance is greater than $1/3$:
\begin{equation}
\frac{k+1}{k-1}a_{ii} + \bar{a}_{i*} - \bar{a}_{*i} - \frac{k+1}{k-1}\bar{a}_{**} > 0, 
\end{equation}
where $\bar{a}_{**} = 1/3\sum_i^3{a_{ii}}$  is the average payoff for two players using the same strategies, $\bar{a}_{i*} = 1/3\sum_{j=1}^3{a_{ij}}$ is the average payoff of strategy $i$, and $\bar{a}_{*i} = 1/3\sum_{j=1}^3{a_{ji}}$ is the average payoff of players when playing against strategy $i$. 

We note that this inequality is, in fact, a special case of the more general condition for multiple strategies in structured populations under rare mutations ($\mu \to 0$)~\cite{tarnita2009strategy}. A given strategy $i$ is favored by natural selection if the inequality holds:
\begin{equation}
\sigma_1a_{ii} + \bar{a}_{i*} - \bar{a}_{*i} - \sigma_1\bar{a}_{**} > 0, 
\end{equation}
with the structural coefficient $\sigma_1 = \frac{k+1}{k-1}$ for lattice populations and the number of strategies $n = 3$. 

\begin{figure*}[htbp]
\centering
 \includegraphics[width=0.9\columnwidth]{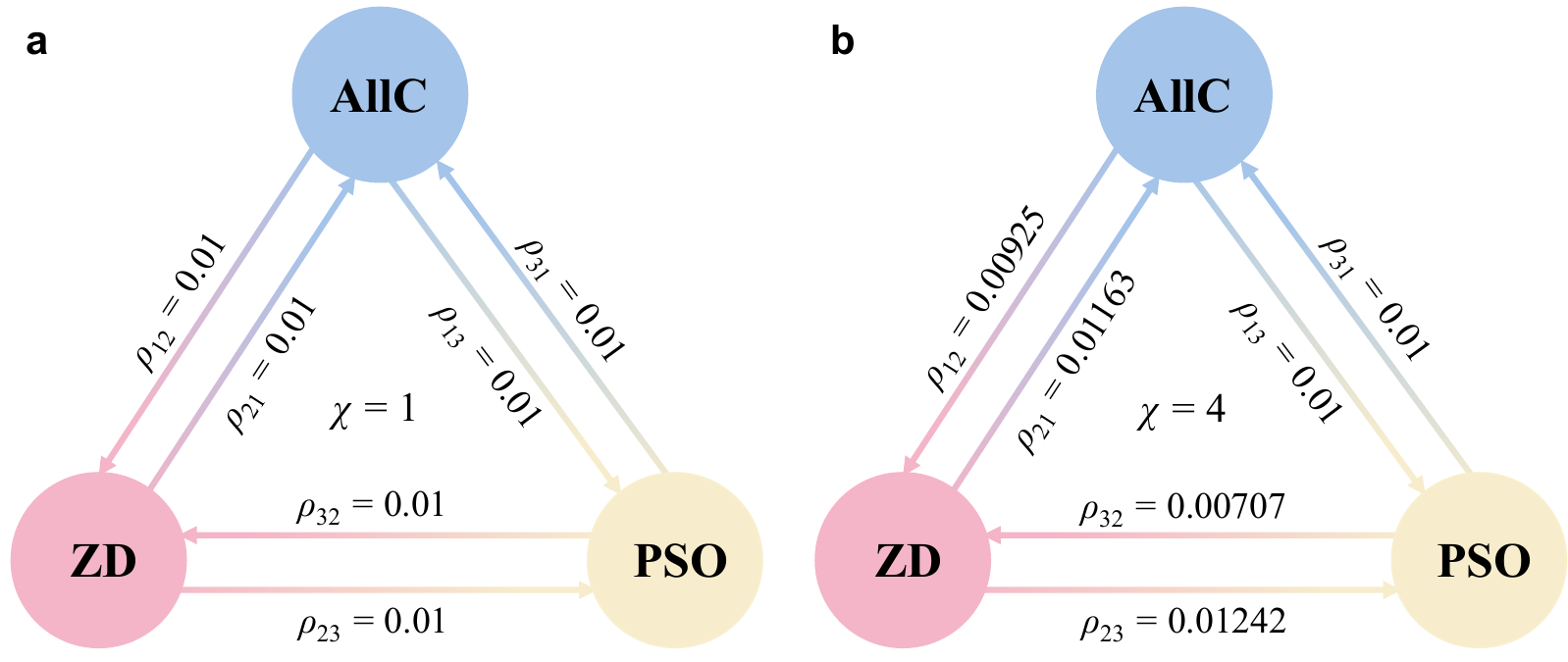}
        \caption{Pairwise competition dynamics under rare mutations. We show fixation probabilities $\rho_{ij}$ between the three strategies considered in repeated games: AllC, extortionate ZD (ZD extortioner) with (a) $\chi = 1$ and (b) $\chi = 4$, and unbending strategy (PSO Gambler), labeled as $i = 1, 2, 3$, respectively. In panel (a), for $\chi = 1$, the game dynamics are neutral among all three strategies, and therefore the fixation probability is $\rho_{ij} = 1/N$ where $N$ is the population size. In panel (b), whereas for $\chi = 4$, the game dynamics remain neutral between AllD and PSO but no longer for AllC vs ZD or ZD vs PSO. It is noteworthy that spatial structure can help AllC to invade extortionate ZD ($\rho_{21} > 1/N > \rho_{12}$) as opposed to well-mixed populations; furthermore, compared to AllC, PSO is less likely to be invaded by ZD and more likely to take over ZD. Together, extortion can be suppressed by spatial structure and the presence of unbending strategies such as the PSO Gambler. Simulation parameters are as in Fig.~1.}
        \label{fig3}
\end{figure*}

When the extortion factor $\chi = 1$, the resulting game dynamics are neutral, with fixation probabilities $\rho_{ij} = 1/N$ for any pairs of $i, j$ (Fig.~3a). Therefore, each strategy has an equal frequency in the long run in the limit of low mutation. However, the neutrality no longer holds for $\chi > 1$ between AllC and extortionate ZD, as well as between PSO Gambler and extortionate ZD, except for AllC and PSO Gambler.

Even though extortionate ZD can now secure the highest payoffs when interacting with AllC or PSO Gambler, it is unable to exploit the unbending PSO Gambler to the same extent as AllC because unbending leads to a monotonic decrease in extortionate ZD's payoff with respect to $\chi$. This makes PSO Gambler harder to invade by extortion ZD and meanwhile easier to take over extortionate ZD, as compared to AllC (Fig.~2). Altogether, the presence of spatial structure can help natural selection favor AllC over extortion because of spatial assortment, and extortioners fail to do well when against each other since they get neutralized and together receive the payoff $P$. Moreover, unbending PSO Gambler, compared to AllC, can even further diminish the advantage of extortion, if any, in spatial populations. 

\begin{figure*}[htbp]
\centering
 \includegraphics[width=0.9\columnwidth]{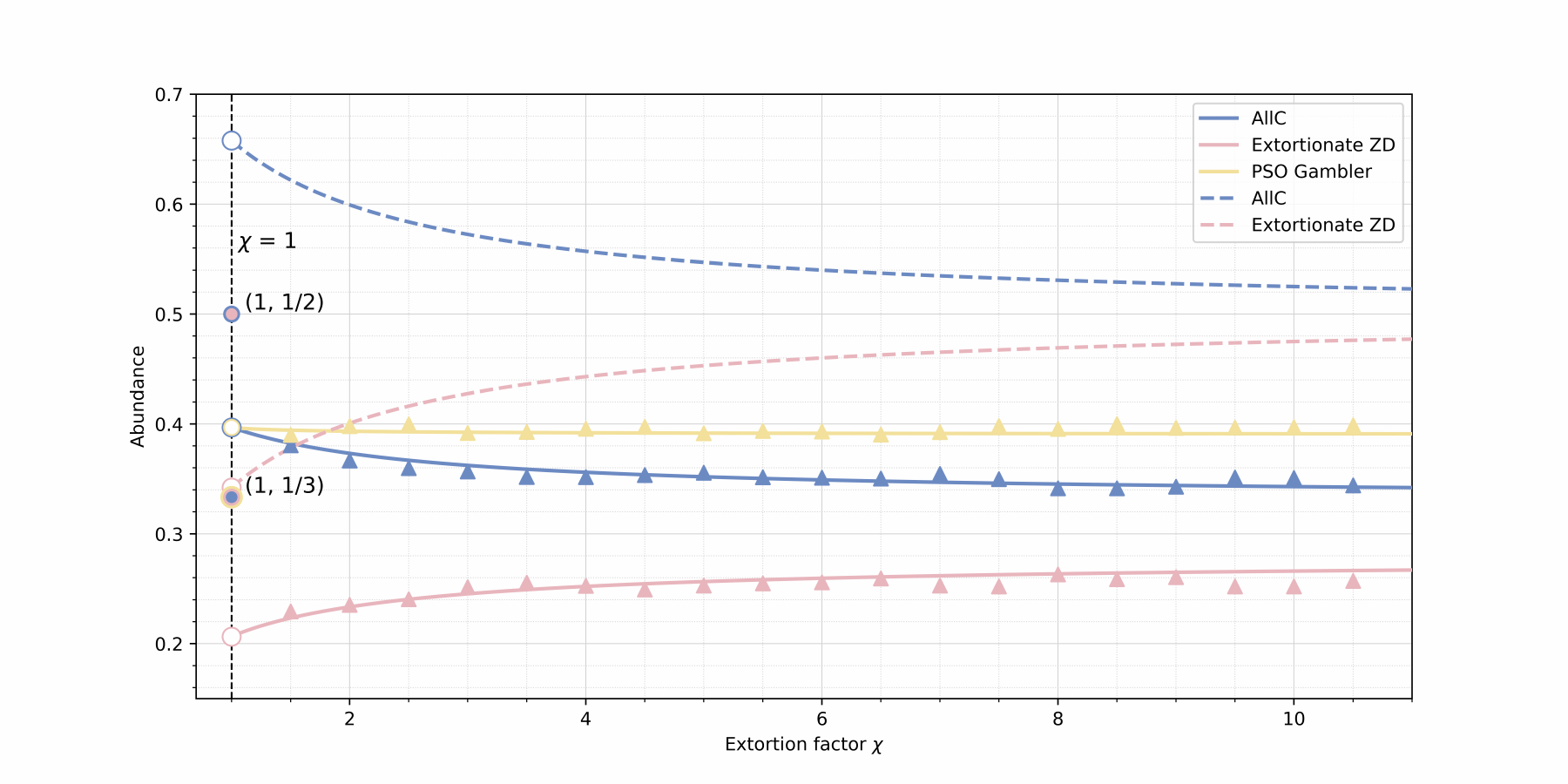}
        \caption{Long-term frequencies of the three strategies -- AllC, extortionate ZD (extortioner), and unbending strategy (PSO Gambler) -- in lattice populations. We find good agreement between the simulation results (symbols) and analytical predictions (lines), shown as a function of the extortion factor $\chi$. Notably, the frequency of PSO Gambler remains almost unchanged while the equilibrium frequency of extortioners increases at the expense of AllC. In reference to the competition of AllC vs ZD in spatial populations (dashed lines), ZD can achieve a half-half split in equilibrium with AllC for sufficiently large extortion factors, despite the role of spatial structure in favoring AllC. However, the presence of the unbending strategy, PSO Gambler, fundamentally affects the underlying pairwise competition dynamics (as depicted in Fig.~3), thereby suppressing extortion regardless of how large $\chi$ is. Simulation parameters are as in Fig.~1., except that we vary the extortion factor $\chi$.}
        \label{fig4}
\end{figure*}
%almost

As demonstrated in Fig.~4, an increase in the extortion factor helps extortionate ZD players increase in abundance as more benefits are squeezed from AllC, but their extortion is greatly mitigated by the presence of an unbending strategy, the PSO Gambler for example. These PSO Gamblers are able to make extortion backfire: the greater $\chi$, the less absolute payoff extortionate ZD can reap from unbending players. Therefore, the presence of a third type of unbending greatly increases the system's resilience against extortion. The abundance of PSO Gamblers is only slightly impacted by increases in $\chi$, quickly reaching a plateau -- in other words, it saturates much faster than the impacts on AllC and extortionate ZD. For the limit $\chi \to \infty$, the abundance of extortionate ZD reaches a limit that is strictly below $1/3$. Our simulation results agree well with the analytical predictions (Fig.~4). This is in sharp contrast with the scenario where PSO Gambler is absent and extortionate ZD can achieve a half-half split in abundance with AllC at the limit $\chi \to \infty$ (see Appendix). Taken together, unbending strategies such as the PSO Gambler can help shepherd cooperation and suppress extortion in spatial populations.

We also note the jump discontinuity of the long-term abundance at $\chi = 1$ versus $\chi > 1$. This is due to the fact that the stochastic memory-one ZD strategy reaches the boundary deterministic strategy Tit-for-Tat, thereby causing the pairwise moves CC to become the only attracting state, as opposed to being ergodic among all four possible pairwise states CC, CD, DC, and DD.

\section*{Discussions and Conclusion}

As a powerful memory-one strategy, an extortionate zero-determinant player can dominate any other co-player (or tie at worst) in the conventional Prisoner's Dilemma game. A recent study reveals that unbending players, when fixed, can render the best response of an extortionate ZD player to be fair by letting their extortion factor approach one~\cite{chen2023outlearning}. Moreover, unbending players can dominate ZD players even if the underlying game remains a Prisoner's Dilemma game but is of a more adversarial nature featuring $T+S < 2P$. In this work, we deepen our understanding of unbending strategies by showing their capacity to enhance spatial cooperation while suppressing extortion. The intuition is that unbending strategies are neutral with AllC while they reduce the payoffs of extortionate ZD strategies due to their unbending properties. Therefore, the presence of unbending individuals can drastically change the invasion dynamics among them under the mutation-selection equilibrium.

Among the body of previous research, certain work has demonstrated that extortionate ZD players cannot be evolutionary successful unless they become more generous~\cite{stewart2013extortion}. These studies typically have been done in well-mixed populations with multiple strategies~\cite{hilbe2013evolution}. It is evident that population size has an effect on the evolutionary advantage of extortion. On the one hand, extortion can dominate in small-sized populations~\cite{hilbe2013evolution}. On the other hand, population structure promotes assortment (termed as network or spatial reciprocity~\cite{nowak1992evolutionary, nowak2006five}), which can further strengthen the advantage of cooperation. Depending on the payoff structure parameters, such as the benefit-to-cost ratio in donation games or other parameterizations of the Prisoner's Dilemma, increasing the extortion factor $\chi$ can help ZD players dominate other strategies and subsequently provide an evolutionary pathway to cooperation, thereby acting as a catalyst for the evolution of cooperation. These insights are obtained in both well-mixed and structured populations~\cite{hilbe2013evolution, szolnoki2014evolution}. Here, the contribution of our current work lies in demonstrating how unbending players can further foster direct reciprocity and suppress extortion, thus increasing the resilience of cooperation against extortion.

In addition to fixation probabilities, another important quantity of interest is the conditional fixation time~\cite{antal2006fixation}. Prior work has shown that fixation time strongly depends on the type of game interactions (payoff structure in general) in finite, well-mixed populations~\cite{antal2006fixation, altrock2009fixation}. For instance, the fixation time is exponential when the underlying game is of the snowdrift type~\cite{antal2006fixation}. Moreover, spatial structure can promote exceedingly long co-existence even if the underlying game is of the Prisoner's Dilemma type~\cite{nowak1992evolutionary}. In our current study, we have a snowdrift game both between AllC and extortionate ZD, and between extortionate ZD and PSO Gambler, and a neutral game between AllC and PSO Gambler. Thus, the fixation time can be prohibitively long for non-weak selection strength in well-mixed populations, not to mention the role of spatial structure in promoting coexistence~\cite{hauert2004spatial}. That said, our analytical approach based on pairwise invasion dynamics can no longer be employed because fixation takes exceedingly long and, as a consequence, we can only observe the co-existence of AllC, extortionate ZD, and PSO Gambler while one strategy being fixed in the population becomes an extremely rare event. In this case, we are compelled to rely on agent-based simulations and extended pair approximation methods for multiple strategies to understand the dynamics. Previous studies utilizing simulations have focused on non-weak selection and offered some insights into this regime~\cite{szolnoki2014evolution, szolnoki2014defection, wu2014boosting, hao2015extortion, xu2017extortion, mao2018emergence}.
% becomes an extremely rare event in simulations

In conclusion, we have analytically and by means of agent-based simulations investigated how introducing unbending strategies can help suppress extortion and shepherd cooperation in spatial populations. We have derived closed-form approximations for the long-term frequencies of three strategies, AllC, extortionate ZD, and the unbending strategy PSO Gambler, under weak selection and in the limit of low mutation. We find that the presence of unbending strategies can restrain the abundance of extortion ZD no matter how large the extortion factor is, whereas the impact of the extortion factor has little effect on the long-run abundance of unbending strategies. Therefore, unbending individuals can strengthen the resilience of spatial cooperation. Although we demonstrate our general method through a particular candidate of unbending strategy, the PSO Gambler, our approach can apply to study broader contexts such as the evolutionary dynamics of multiple powerful strategies in the repeated multiplayer games~\cite{hilbe2014cooperation, pan2015zero, szolnoki2018environmental, shao2019evolutionary, wang2020steering}, multiplex networks~\cite{battiston2017determinants}, higher order networks~\cite{alvarez2021evolutionary}, and enforcing fairness in human-AI systems~\cite{santos2019evolution, wang2024mathematics}, providing insights into understanding the interplay between network reciprocity and direct reciprocity.

\ack

We would like to express our heartfelt gratitude to Professor Long Wang on the occasion of his 60th birthday. X.C. gratefully acknowledges the support by Beijing Natural Science Foundation (grant no.~1244045). F.F. is grateful for support from the Bill \& Melinda Gates Foundation (award no.~OPP1217336).

\section*{Author Contributions} 

Z.C., X.C., and F.F. conceived the model; Z.C. and Y.G. performed calculations and analyses and plotted the figures; Z.C., X.C., and F.F. wrote the manuscript. All authors give final approval of publication.

\section*{Competing Interests} 

The authors declare that they have no competing financial interests.

\appendix

\section{}

In our study, we consider a population of size $N$ under a network structure of degree $k$ where individuals can be of strategy A or B. The corresponding payoff matrix between these two strategies is $(a, b, c, d)$. For the evolution process, the death-birth update rule is used and the contribution of payoff to fitness is described by the selection strength $\beta$. We first work on pairwise invasions with no mutations ($\mu = 0$) and then consider the evolutionary dynamics of a triple of strategies under rare mutations ($\mu \to 0$).

\subsection{Fixation probabilities}

The probability of randomly picking an individual of type X is denoted by $p_{\rm{X}}$. Moreover, the probability of randomly picking an XY pair is denoted by $p_{\rm{XY}}$ and the conditional probability of finding a neighbor of type X given a focal individual of type Y is referred to as $q_{\rm{X|Y}}$. It follows immediately that $p_{\rm{XY}} = p_{\rm{Y}}q_{\rm{X|Y}}$. We also have $p_A + p_B = 1$, $p_A = p_{AA} + p_{BA}$, $p_B = p_{AB} + p_{BB}$, and $p_{AB} = p_{BA}$. Therefore, the two probabilities $p_A$ and $p_{AA}$ can be considered as the independent variables of the population. In the limit of weak selection where $\beta \ll 1$, it has been proven in previous work that we can further obtain an elegant relation between the global frequency $p_{\rm{X}}$ and the local density $q_{\rm{X|X}}$ via the degree $k$. To be more precise, the following two identities hold:
\begin{equation}
q_{A|A} = p_A + \frac{1 - p_A}{k - 1}, \qquad q_{B|B} = p_B + \frac{1 - p_B}{k - 1}.
\label{manifold}
\end{equation}

We assume that there are $i$ individuals of strategy A and $N - i$ individuals of strategy B at the moment. If a focal A individual is selected, let the numbers of its neighbors of type A and type B be $n_A^A(i)$ and $n_A^B(i)$, respectively. On average, we have $n_A^A(i) = kq_{A|A}$ and $n_A^B(i) = kq_{B|A}$. The payoffs of these two types of neighbors are
\begin{equation}
\eqalign{
\pi_{A}^{A}(i) &= \left[(k-1)q_{A|A}+1\right]\cdot a+(k-1)q_{B|A}\cdot b, \\
\pi_{A}^{B}(i) &= \left[(k-1)q_{A|B}+1\right]\cdot c+(k-1)q_{B|B}\cdot d.
}
\label{payoff_A}
\end{equation}
Similarly, if a focal B individual is selected, let the numbers of its neighbors of type A and type B be $n_B^A(i)$ and $n_B^B(i)$, respectively. On average, we have $n_B^A(i) = kq_{A|B}$ and $n_B^B(i) = kq_{B|B}$. The payoffs of these two types of neighbors are
\begin{equation}
\eqalign{
\pi_{B}^{A}(i) &= (k-1)q_{A|A}\cdot a+\left[(k-1)q_{B|A}+1\right]\cdot b, \\
\pi_{B}^{B}(i) &= (k-1)q_{A|B}\cdot c+\left[(k-1)q_{B|B}+1\right]\cdot d,
}
\label{payoff_B}
\end{equation}
The corresponding fitness can be written as 
\begin{equation}
\eqalign{
f_A^A(i) &= \exp[\beta\pi_A^A(i)], \qquad
f_A^B(i) = \exp[\beta\pi_A^B(i)]. \\
f_B^A(i) &= \exp[\beta\pi_B^A(i)], \qquad
f_B^B(i) = \exp[\beta\pi_B^B(i)].
}
\label{fitness}
\end{equation}

Based on the update rule, the probability that the number of A individuals decreases by one is
\begin{equation}
T_{A}^{-}(i) = p_{A}\frac{n_{A}^{B}(i)f_{A}^{B}(i)}{n_{A}^{A}(i)f_{A}^{A}(i)+n_{A}^{B}(i)f_{A}^{B}(i)},
\end{equation}
whereas the probability that the number of A individuals increases by one is
\begin{equation}
T_{A}^{+}(i) = p_{B}\frac{n_{B}^{A}(i)f_{B}^{A}(i)}{n_{B}^{A}(i)f_{B}^{A}(i)+n_{B}^{B}(i)f_{B}^{B}(i)}.
\end{equation}
The evolution process can be regarded as a finite-state Markov chain in the interval $[0, N]$, where the endpoints $0$ and $N$ are the absorbing states. After some routine calculations, we can obtain the fixation probability for strategy A:
\begin{equation}
\rho_{A} = \frac{1}{1+\sum_{i=1}^{N-1}\prod_{j=1}^{i}\frac{T_{A}^{-}(j)}{T_{A}^{+}(j)}}.
\label{rho}
\end{equation}
We can further get the condition for natural selection to favor strategy A:
\begin{equation}
\rho_A > \frac{1}{N} \Leftrightarrow 1+\sum_{i=1}^{N-1}\prod_{j=1}^i\frac{T_A^-(j)}{T_A^+(j)} < N.
\label{favor_00}
\end{equation}

We know that $T_{A}^{-}(i) = T_{A}^{+}(i) = p_{AB}$ when the selection strength $\beta = 0$. In the limit of weak selection where $\beta \ll 1$, we can take the Taylor expansion of the ratio $T_{A}^{-}(i)/T_{A}^{+}(i)$ to the first order and get
\begin{equation}
\frac{T_{A}^{-}(j)}{T_{A}^{+}(j)} = 1+\beta\big\{q_{A|A}[\pi_{A}^{B}(i)-\pi_{A}^{A}(i)]+q_{B|B}[\pi_{B}^{B}(i)-\pi_{B}^{A}(i)]\big\}+\mathcal{O}(\beta^2).
\end{equation}
As such, the condition in \ref{favor_00} is simplified as
\begin{equation}
\sum_{i=1}^{N-1} \sum_{j=1}^{i} \big\{q_{A|A}[\pi_{A}^{A}(i)-\pi_{A}^{B}(i)]+q_{B|B}[\pi_{B}^{A}(i)-\pi_{B}^{B}(i)]\big\} > 0.
\end{equation}
After substituting \ref{manifold} - \ref{fitness} into the above inequality, it becomes
\begin{equation}
\sum_{i=1}^{N-1} \sum_{j=1}^{i} k(F_{a}+F_{b}-F_{c}-F_{d}) > 0,
\label{favor_0}
\end{equation}
where
\begin{equation}
\eqalign{
F_{a} &= {\frac{p_{A}(k^{2}-k-2)+(k+1)}{k(k-1)}} \cdot a, \\
F_{b} &= {\frac{p_{A}(-k^{2}+k+2)+(k^{2}-k-1)}{k(k-1)}} \cdot b, \\
F_{c} &= {\frac{p_{A}(k^{2}-k-2)+1}{k(k-1)}} \cdot c, \\
F_{d} &= {\frac{p_{A}(-k^{2}+k+2)+(k^{2}-1)}{k(k-1)}} \cdot d.
}
\end{equation}

If the population size is large enough, that is, $N \gg k$, a discrete sum can be approximated by a continuous integral. Therefore, we can replace discrete variables $i, j \in \{0, 1, 2, \cdots, N\}$ by continuous variables $u, v \in (0, 1)$. And the condition in \ref{favor_0} is simplified as
\begin{equation}
 \int_{0}^{1} \int_{0}^{u} (H_{a}+H_{b}-H_{c}-H_{d}) dvdu > 0,
\end{equation}
where
\begin{equation}
\eqalign{
H_{a} &= [v(k^{2}-k-2)+(k+1)] \cdot a,  \\
H_{b} &= [v(-k^{2}+k+2)+(k^{2}-k-1)] \cdot b, \\
H_{c} &= [v(k^{2}-k-2)+1] \cdot c, \\
H_{d} &= [v(-k^{2}+k+2)+(k^{2}-1)] \cdot d,
}
\end{equation}
By integrating the four parts, we finally get the condition for natural selection to favor strategy A, determined by the payoff structure and the network structure:
\begin{equation}
(k + 1)^2a + (2k^2 - 2k - 1)b - (k^2 - k + 1)c - (2k - 1)(k + 1)d.
\label{favor}
\end{equation}

In like manner, we obtain the fixation probabilities for both strategies as linear functions of the selection strength:
\begin{equation}
\fl
\eqalign{
\rho_{A} = \frac{1}{N}+\beta \cdot \frac{(k+1)^{2}a+(2k^{2}-2k-1)b-(k^{2}-k+1)c-(2k-1)(k+1)d}{6(k-1)}, \\
\rho_{B} = \frac{1}{N}+\beta \cdot \frac{(k+1)^{2}d+(2k^{2}-2k-1)c-(k^{2}-k+1)b-(2k-1)(k+1)a}{6(k-1)}.
}
\end{equation}

\subsection{Average payoffs of IPD strategies}

We consider the conventional IPD game with payoff matrix $(R, S, T, P) = (3, 0, 5, 1)$.  Then, we focus on three strategies: AllC (cooperator), extortionate ZD (extortioner), and unbending strategy (PSO Gambler). Since these strategies are memory-one, each of them can be represented by a four-component vector $\bm{q} = [q_{\rm{CC}}, q_{\rm{CD}}, q_{\rm{DC}}, q_{\rm{DD}}]$. Here, $q_{\rm{CC}}$ is the probability of cooperating in the next round given the outcome CC in the current round and the same applies to the other components. We denote the vectors of AllC, extortionate ZD, and PSO Gambler by $\bm{q}_1$, $\bm{q}_2$, and $\bm{q}_3$, respectively.

To be more precise, we have
\begin{equation}
\footnotesize{
\fl
   \mathbf{q}_{\rm{AllC}} = \left[
    \begin{array}{c}
    1 \\
    1 \\
    1 \\
    1
    \end{array}
    \right],
    \quad
    \mathbf{q}_{\rm{ZD}}  = \left[
    \begin{array}{c}
        1 - \phi(R - P)(\chi - 1) \\
        1 - \phi[(T - P)\chi + (P - S)] \\
        \phi[(P - S)\chi + (T - P)] \\
        0
    \end{array}
    \right],
    \quad
   \mathbf{q}_{\rm{PSO}}  = \left[
    \begin{array}{c}
    1 \\
    0.52173487 \\
    0 \\
    0.12050939
    \end{array}
    \right],
}
\end{equation}
where $\chi \geq 1$ is the extortion factor and $0 \leq \phi \leq (P - S)/[(T - P)\chi + (P - S)]$ is the normalization factor. Although $\phi$ has a non-trivial effect on the payoffs between extortionate ZD and the co-player in general, the two strategies discussed here can actually trivialize the influence enforced by $\phi$.

We use the method (taking the quotient of the determinants of two matrices) introduced by Press and Dyson in their work~\cite{press2012iterated} to calculate the expected payoffs $s_{\rm{X}}$ and $s_{\rm{Y}}$ between a pair of players X and Y using strategy $\bm{p}$ and $\bm{q}$, respectively:
\begin{equation}
s_{\rm{X}} = \frac{D(\bm{p},\bm{q},\bm{S}_{\rm{X}})}{D(\bm{p},\bm{q},\bm{I})}, \qquad
s_{\rm{Y}} = \frac{D(\bm{p},\bm{q},\bm{S}_{\rm{Y}})}{D(\bm{p},\bm{q},\bm{I})},
\end{equation}
where $\bm{S}_{\rm{X}} = (R, S, T, P)$,  $\bm{S}_{\rm{Y}} = (R, T, S, P)$, and $\bm{I} = (1, 1, 1, 1)$. We refer to the $3 \times 3$ expected payoff matrix of AllC, extortionate ZD, and PSO Gambler as
\begin{equation}
\begin{blockarray}{cccc}
& \rm{AllC} & \rm{ZD} & \rm{PSO} \\
\begin{block}{c[ccc]}
\rm{AllC}\,\,\, & a_{11} & a_{12} & a_{13} \\
\rm{ZD}\,\,\, & a_{21} & a_{22} & a_{23} \\
\rm{PSO}\,\,\, & a_{31} & a_{32} & a_{33} \\
\end{block}
\end{blockarray}
\,\,\, .
\end{equation}

It always holds that $a_{11} = a_{13} = a_{31} = a_{33} = R$. When $\chi = 1$, it follows that $\bm{q}_2 = [1, 0, 1, 0]$, namely, extortionate ZD degenerates to Tit-for-Tat. We further have $a_{11} = a_{12} = a_{21} = a_{22} = a_{23} = a_{32} = R$. In contrast, when $\chi > 1$, we have
\begin{equation}
\eqalign{
a_{11} &= R,\qquad a_{12} = P+\frac{(R-P)(T-S)}{(R-S)\chi+(T-R)}, \\
a_{22} &= P, \qquad a_{21} = P+\frac{(R-P)(T-S)\chi}{(R-S)\chi+(T-R)}.
}
\end{equation}
And the remaining two payoffs $a_{23}$ and $a_{32}$ become quadratic rational functions of $\chi$. In the conventional IPD game, the four payoffs decided by $\chi$ can be (approximately) written as
\begin{equation}
\eqalign{
a_{12} &= \frac{3(\chi+4)}{3\chi+2},  \qquad a_{21} = \frac{13\chi+2}{3\chi+2}, \\
a_{23} &= \frac{0.455742281814226(\chi-0.598718917613494)(4\chi+1)}{\chi^{2}-0.422337214095301\chi-0.272861525678518}, \\
a_{32} &= \frac{\chi^{2}+0.400631913161605\chi-0.48622813248306}{\chi^{2}-0.422337214095301\chi-0.272861525678518}.
}
\end{equation}
Moreover, we can get the extreme values of these payoffs by letting $\chi \to +\infty$:
\begin{equation}
\fl
\lim_{\chi \to +\infty}a_{12} = \lim_{\chi \to +\infty}a_{32} = 1, \quad \lim_{\chi \to +\infty}a_{21} = \frac{13}{3}, \quad
\lim_{\chi \to +\infty}a_{23} = 1.82296912725691.
\end{equation}

\subsection{Impact of extortion factor on long-term frequencies of IPD strategies}

In a similar manner, we denote the pairwise fixation probabilities between AllC (strategy $1$), extortionate ZD (strategy $2$), and PSO Gambler (strategy $3$) by $\rho_{ij}$'s and the long-term frequencies of these strategies by $\tilde{\lambda}_i$'s when unbending Gamblers are absent from the population (the null case) or $\lambda_i$'s when all three strategies are present. 

Based on the work of  Fudenberg and Imhof~\cite{fudenberg2006imitation}, we use the embedded Markov chain approach and obtain these frequencies as elements of the normalized left eigenvector of the transition matrix corresponding to the largest eigenvalue one. For two strategies, we have
\begin{equation}
[\tilde{\lambda}_1, \tilde{\lambda}_2] =\frac{1}{\rho_{21} + \rho_{12}}[\rho_{21}, \rho_{12}].
\label{frequency2}
\end{equation}
And for three strategies, we have
\begin{equation}
[\lambda_1, \lambda_2, \lambda_3] =\frac{1}{\gamma_1 + \gamma_2 +\gamma_3}[\gamma_1, \gamma_2, \gamma_3],
\label{frequency3}
\end{equation}
where 
\begin{equation}
\eqalign{
\gamma_1 &= \rho_{21}\rho_{31} + \rho_{21}\rho_{32} + \rho_{31}\rho_{23}, \\
\gamma_2 &= \rho_{31}\rho_{12} + \rho_{12}\rho_{32} + \rho_{32}\rho_{13}, \\
\gamma_3 &= \rho_{21}\rho_{13} + \rho_{12}\rho_{23} + \rho_{13}\rho_{23}.
}
\label{frequency3X}
\end{equation}

\begin{figure*}[htbp]
\centering
 \includegraphics[width=0.9\columnwidth]{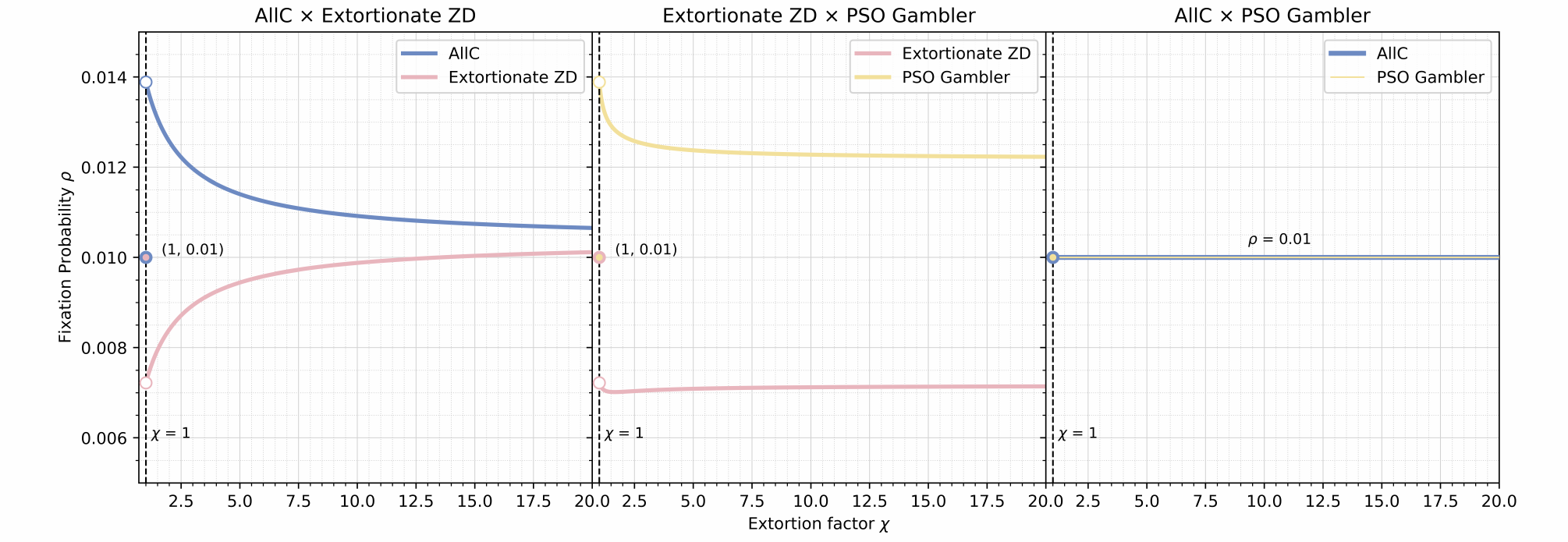}
        \caption{Fixation probabilities between the three strategies considered in repeated games: AllC, extortionate zero-determinant strategy (ZD extortioner) with $\chi \geq 1$, and unbending strategy (PSO Gambler). Simulation parameters are as in Fig.~1, except that we vary the extortion factor $\chi$ for the extortionate ZD strategy.}
        \label{figA1}
\end{figure*}

To study the impact of the extortion factor $\chi$ on $\rho_{ij}$ and $\lambda_i$, we consider a specific example where $N = 100$, $k = 4$, $\beta = 0.001$ to get concrete numerical results. The same approach applies to any other combination of parameters as long as the conditions $N \gg k$ and $\beta \ll 1$ are satisfied. After some tedious calculations, we get
\begin{itemize}
\item when $\chi = 1$, 
\begin{equation}
\rho_{21} = \rho_{12} = \rho_{31} = \rho_{23} = \rho_{31} = \rho_{13} = \frac{1}{100},
\end{equation}
\item when $\chi > 1$,
\begin{equation}
\fl
\eqalign{
\rho_{21} &= \frac{28\chi+34.5}{900(3\chi+2)}, \quad \rho_{12} = \frac{28\chi+4.5}{900(3\chi+2)}, \\
\rho_{32} &= \frac{2.14880483211515\chi^{2}-1.03438540268879\chi-0.454016698936298}{300(\chi^{2}-0.422337214095301\chi-0.272861525678518)} \\
\rho_{23} &= \frac{3.655023355761\chi^{2}+1.25725839044252\chi+1.12775971437606}{300(\chi^{2}-0.422337214095301\chi-0.272861525678518)}, \\
\rho_{31} &= \rho_{13} = \frac{1}{100}.
}
\end{equation}
\end{itemize}
It is clear that these fixation probabilities can be seen as functions of the extortion factor. The curves of $\rho_{ij}$'s with respect to $\chi$ are given in Fig~\ref{figA1}.

\begin{figure*}[htbp]
\centering
\includegraphics[width=0.9\columnwidth]{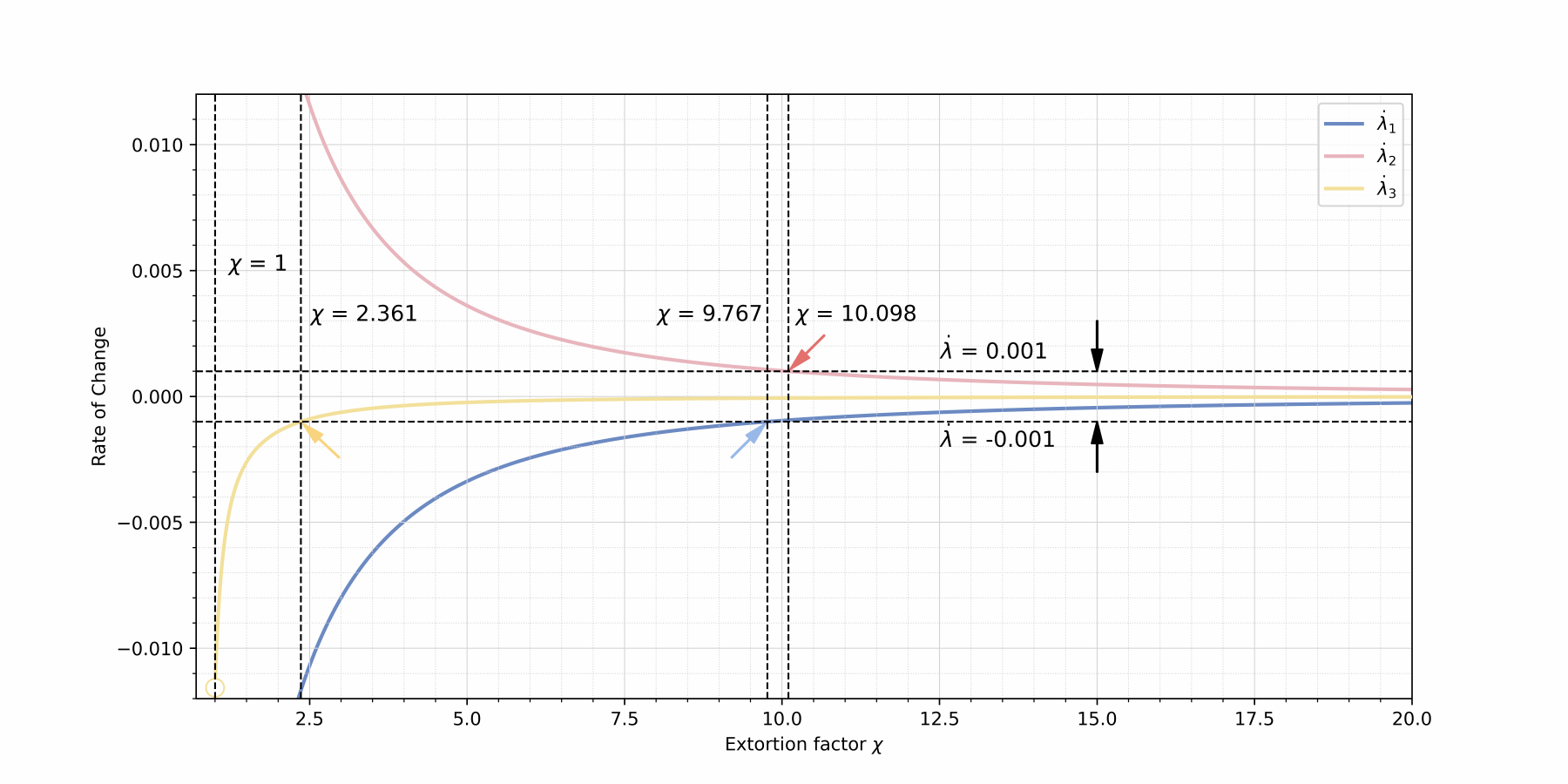}
        \caption{Derivatives of long-term frequencies of the three strategies -- AllC, extortionate ZD, and PSO Gambler -- in lattice populations. Simulation parameters are as in Fig.~1, except that we vary the extortion factor $\chi$.}
        \label{figA2}
\end{figure*}

Combined with \ref{frequency2} - \ref{frequency3X}, we further get
\begin{itemize}
\item when $\chi = 1$, $[\tilde{\lambda}_1, \tilde{\lambda}_2] = [1/2, 1/2]$ and $[\lambda_1, \lambda_2, \lambda_3] = [1/3, 1/3, 1/3]$,
\item when $\chi > 1$,
\begin{equation}
[\tilde{\lambda}_1, \tilde{\lambda}_2] = [\frac{28\chi+34.5}{56\chi+39}, \frac{28\chi+4.5}{56\chi+39}],
\end{equation}
and
\begin{equation}
\fl
\scriptsize{
\eqalign{
\lambda_{1} &= \frac{1}{3} \cdot \frac{4.49726233156984\chi^{3} + 2.68590913537835\chi^{2} - 3.11317031657329\chi - 1.18897071999626}{4.50655997677753\chi^{3} + 1.20760596881118\chi^{2} - 2.48101898478848\chi - 0.813081399282176}, \\
\lambda_{2} &= \frac{1}{3} \cdot \frac{3.74415306974691\chi^{3} - 0.565174340912924\chi^{2} - 1.42345389906067\chi - 0.25738900597642}{4.50655997677753\chi^{3} + 1.20760596881118\chi^{2} - 2.48101898478848\chi - 0.813081399282176} \\
\lambda_{3} &= \frac{1}{3} \cdot \frac{5.27826452901584\chi^{3} + 1.50208311196812\chi^{2} - 2.90643273873147\chi - 0.9928844718738\chi}{4.50655997677753\chi^{3} + 1.20760596881118\chi^{2} - 2.48101898478848\chi - 0.813081399282176}. 
}
}
\end{equation}
\end{itemize}
As before, these abundances can be seen as functions of the extortion factor. The curves of $\tilde{\lambda}_i$'s and $\lambda_i$'s with respect to $\chi$ are given in Fig~\ref{fig4}. In particular, we can get the extreme values of these abundances by letting $\chi \to +\infty$:
\begin{equation}
\eqalign{
\lim_{\chi \to +\infty}\tilde{\lambda}_{1} &= \lim_{\chi \to +\infty}\tilde{\lambda}_{2}=\frac{1}{2}, \\
\lim_{\chi \to +\infty}\lambda_{1} &= 0.332645621401128, \\
\lim_{\chi \to +\infty}\lambda_{2} &= 0.276940954892473, \\
\lim_{\chi \to +\infty}\lambda_{3} &= 0.390413423706399, 
}
\end{equation}
It can be observed from Fig~\ref{fig4} that the abundance of PSO Gambler is only slightly impacted by increases in the extortion factor, compared with those of AllC and extortionate ZD. To understand the trends in a more intuitive way, we differentiate $\lambda_i$'s with respective to $\chi$ and get their corresponding derivatives. The curves of $\dot{\lambda}_i$'s are shown in Fig~\ref{figA2}.

\section{}

Data Availability: All the data and analyses pertaining to this work have been included in the main text. \\
Code Availability: The source code for reproducing the results is available at the GitHub repository (https://github.com/fufeng/unbending3S).

\end{document}